\documentclass[aps,twocolumn,twoside,nofootinbib,preprintnumbers]{revtex4}

\usepackage{graphicx,amsmath,hyperref}
\newcommand{\<}{\langle}
\renewcommand{\>}{\rangle}

\newcommand{\be}{\begin{equation}}
\newcommand{\ee}{\end{equation}}
\def\ba#1\ea{\begin{align}#1\end{align}} 

\newcommand{\cond}[1]{\left\{\begin{array}{l@{~~~}l}#1\end{array}\right.}

\newcommand{\E}{{\cal E}}
\newcommand{\Hfree}{H_0}
\renewcommand{\d}{{\mathrm{d}}}
\renewcommand{\k}{k}
\newcommand{\p}{p}
\newcommand{\x}{x}
\newcommand{\ej}{e_j}

\begin{document}

\title{Spatial search and the Dirac equation}

\author{Andrew M. Childs}
\email[]{amchilds@mit.edu}

\affiliation{Center for Theoretical Physics,
             Massachusetts Institute of Technology,
             Cambridge, MA 02139, USA}

\author{Jeffrey Goldstone}
\email[]{goldston@mit.edu}

\affiliation{Center for Theoretical Physics,
             Massachusetts Institute of Technology,
             Cambridge, MA 02139, USA}

\date[]{20 May 2004}

\preprint{MIT-CTP \#3495}


\begin{abstract}
We consider the problem of searching a $d$-dimensional lattice of $N$
sites for a single marked location.  We present a Hamiltonian that
solves this problem in time of order $\sqrt N$ for $d>2$ and of order
$\sqrt N \log N$ in the critical dimension $d=2$.  This improves upon
the performance of our previous quantum walk search algorithm (which
has a critical dimension of $d=4$), and matches the performance of a
corresponding discrete-time quantum walk algorithm.  The improvement
uses a lattice version of the Dirac Hamiltonian, and thus requires the
introduction of spin (or coin) degrees of freedom.
\end{abstract}


\maketitle

\section{Introduction}

Quantum mechanical computers can solve certain problems asymptotically
faster than classical computers.  One of the major advantages of
quantum computation comes from a fast algorithm for the problem of
finding a marked item among $N$ items.  Whereas a classical computer
requires $\Theta(N)$ steps to solve this problem, Grover showed that a
quantum computer can solve it using only $O(\sqrt N)$ steps
\cite{Gro97}, which is optimal \cite{BBBV97}.

To apply Grover's algorithm, it must be possible to quickly perform a
reflection about a superposition of all possible items.  However, this
may not be feasible if the items are distributed in space and the
algorithm is restricted to access them by local moves.  For example,
if the items are arranged on a one-dimensional line, simply traveling
from one end of the line to the other requires $N$ moves, and a
straightforward argument shows that no local algorithm, classical or
quantum, can find a marked item in less time than $\Omega(N)$.  But
for other geometries, such as higher dimensional lattices, a quantum
algorithm can conceivably achieve a speedup over the classical
complexity of $\Theta(N)$.

Recently, there has been considerable progress in understanding the
spatial search problem for quantum computers.  Aaronson and Ambainis
gave an algorithm that finds a marked item in the optimal time
$O(\sqrt N)$ for a lattice in $d>2$ dimensions, and in time $O(\sqrt N
\log^2 N)$ for a two-dimensional lattice \cite{AA03}.  Their algorithm
is based on a carefully optimized recursive search of subcubes, which
raises the question of whether a simpler algorithm could solve the
problem just as quickly (or perhaps even faster in two dimensions).
In particular, it is interesting to consider {\em quantum walk}
algorithms, which only use local, time-independent dynamics.  Two
distinct kinds of quantum walk algorithms have been considered.  In
the {\em continuous-time} quantum walk \cite{FG98}, the algorithm is
described by a time-independent Hamiltonian connecting adjacent sites.
In the {\em discrete-time} quantum walk \cite{Wat01a,AAKV01,ABNVW01},
the algorithm consists of repeated application of a fixed local
unitary transformation.

In \cite{CG03}, we considered a continuous-time quantum walk algorithm
for the spatial search problem using no additional memory beyond the
present location of the walker.  We showed that this algorithm can
find a single marked site in time $O(\sqrt N)$ for dimensions $d>4$
and in time $O(\sqrt N \log N)$ in four dimensions.\footnote{The run
time $O(\sqrt N \log N)$ in $d=4$ can be achieved using amplitude
amplification \cite{BHMT00}.  Using only classical repetition of the
quantum walk, the algorithm requires $O(\sqrt N \log^{3/2} N)$ steps.
The same remark applies to the discrete-time quantum walk algorithm in
$d=2$.}  We also showed that this algorithm fails to provide an
interesting speedup for dimensions $d<4$.  Subsequently, Ambainis,
Kempe, and Rivosh found a discrete-time quantum walk algorithm that
works in lower dimensions \cite{AKR04}.  This algorithm runs in time
$O(\sqrt N)$ for $d>2$ and in time $O(\sqrt N \log N)$ in two
dimensions.  Because a discrete-time quantum walk cannot be defined on
a state space consisting only of the vertices of a graph
\cite{Mey96b}, the algorithm of \cite{AKR04} necessarily uses
additional memory (sometimes referred to as a ``coin'' in analogy to
classical random walks).  In this paper, we consider a continuous-time
quantum walk using additional memory, and we show that it achieves the
same running times as the discrete-time algorithm.

Through the analysis of \cite{CG03}, the failure of the
continuous-time quantum walk algorithm for $d<4$ can be viewed as a
consequence of a quadratic dispersion relation: states with small
momentum have energy proportional to their momentum squared.  If the
dispersion were linear instead of quadratic, so that states with small
momentum had energy proportional to their momentum, one might expect
the algorithm to work whenever $d>2$.  A natural way to achieve linear
dispersion is to employ the massless Dirac equation.  Indeed, using an
appropriate lattice version of the Dirac Hamiltonian, we find a fast
algorithm for spatial search in $d>2$.  Because the Dirac particle
necessarily possesses spin degrees of freedom, the resulting algorithm
must have additional memory beyond the present location.  Thus,
although a continuous-time quantum walk can be defined without
additional memory, we find that the additional degrees of freedom can
improve the algorithm's performance.

\section{Hamiltonians for spatial search}

In the Hamiltonian formulation of the spatial search problem, our goal
is to write down a local Hamiltonian that will quickly transform a
simple initial state, such as the uniform superposition over all
lattice sites
\be
  |s\> = \frac{1}{\sqrt N} \sum_\x |\x\>
\,,
\label{eq:uniform}
\ee
to a state with substantial overlap on the marked state $|w\>$.  In
\cite{CG03}, we considered the Hamiltonian 
\be
  H_{\mbox{\footnotesize\cite{CG03}}} = -\gamma L - |w\>\<w|
\,.
\label{eq:oldham}
\ee
Here the second term identifies the marked location, $\gamma$ is an
adjustable parameter, and $L$ is the Laplacian of an $N$-site square
lattice in $d$ dimensions, periodic in each direction with period
$N^{1/d}$.  $L$ has matrix elements
\be
  \<\x'|L|\x\> = \cond{1   & \x \textrm{~adjacent to~} \x' \\
                       -2d & \x = \x' \\
                       0   & \textrm{otherwise.}}
\ee
It is called the Laplacian because it is a discrete approximation to
the continuum operator $\nabla^2$.

Since the free Hamiltonian $-\gamma L$ is translationally invariant, its
eigenstates are the momentum eigenstates
\be
  |\k\> = \frac{1}{\sqrt N} \sum_\x e^{i \k \cdot \x} |\x\>
\ee
where $k_j = \frac{2 \pi m_j}{N^{1/d}}$, with $m_j = 0,\pm
1,\ldots,\pm\frac{1}{2}(N^{1/d}-1)$ for $N^{1/d}$ odd, and $m_j =
0,\pm 1,\ldots,\pm\frac{1}{2}(N^{1/d}-2),+\frac{1}{2}N^{1/d}$ for
$N^{1/d}$ even.  The energy of the state $|\k\>$ is $\E_L(\k) = 2
\gamma(1-\sum_{j=1}^d \cos k_j)$.  Thus, for small $|\k|$, $\E_L(\k)
\approx \gamma\k^2$.  This quadratic dispersion relation ultimately gives
rise to the critical dimension $d=4$ for the algorithm of \cite{CG03}.

To achieve linear dispersion, we can replace $-\nabla^2$ by the
massless Dirac Hamiltonian.  In the general case of mass $m$, this
Hamiltonian has the form \cite{Dir28}
\be
  H_{\rm Dirac} = \sum_{j=1}^d \alpha_j p_j + \beta m
\,,
\label{eq:dirac}
\ee
where the operators $\alpha_j$ and $\beta$ act on spin degrees of
freedom, and $\p = -i \frac{\d}{\d\x}$ is the momentum operator.  If
the spin operators $\alpha_j$ and $\beta$ satisfy the anticommutation
relations
\be
  \{\alpha_j,\alpha_k\} = 2 \delta_{j,k} \,, \quad
  \{\alpha_j,\beta\} = 0 \,, \quad
  \beta^2 = 1
\,,
\label{eq:anticom}
\ee
then one finds $H_{\rm Dirac}^2 = |\p|^2 + m^2$, as required for a
relativistic particle.  Thus for $m=0$, $H_{\rm Dirac}$ has linear
dispersion, $E_{\rm Dirac} = \pm |\p|$.

To write down the Dirac equation in $d$ dimensions, we need $d+1$
anticommuting operators.  The minimal representation of the algebra
(\ref{eq:anticom}) uses $2^{\lceil d/2 \rceil}$-dimensional matrices,
and hence there are $2^{\lceil d/2 \rceil}$ spin components.

Now consider a lattice version of the massless Dirac Hamiltonian,
equation (\ref{eq:dirac}) with $m=0$.  The continuum operator $p_j$
can be discretely approximated as
\be
  P_j |\x\> = \frac{i}{2} (|\x + \ej\> - |\x - \ej\>)
\,,
\label{eq:latticep}
\ee
where $\ej$ is a unit vector in the $j$ direction.  However, as we
will see later, it turns out that simply taking the free Hamiltonian
(\ref{eq:dirac}) using the lattice approximation (\ref{eq:latticep})
is insufficient.  Instead, we will take\footnote{This choice is
closely related to a standard remedy for the fermion doubling problem
in lattice field theory \cite[p.~27]{Cre83}.}
\be
  \Hfree = \omega \sum_j \alpha_j P_j + \gamma \beta L
\label{eq:hfree}
\ee
where both $\omega$ and $\gamma$ are adjustable parameters.  For a
Hamiltonian with spin degrees of freedom, translation invariance shows
that the eigenstates have the form $|\eta,\k\>$, where $|\eta\>$ is a
(momentum-dependent) spin state.  For (\ref{eq:hfree}), we find states
with energies
\be
  \E(\k) = \pm \sqrt{\omega^2 s^2(\k) + \gamma^2 c^2(\k)}
\,,
\label{eq:freespect}
\ee
where
\be
  s^2(\k) = \sum_{j=1}^d \sin^2 k_j \,, \quad
  c(\k) = 2\sum_{j=1}^d (1-\cos k_j)
\,.
\ee
For small momenta, we have $\E(\k) \approx \pm \omega |\k|$, which leads to
a better search algorithm in low dimensions.

The full algorithm is as follows.  We begin in the state $|\eta,s\>$,
where $|\eta\>$ is any spin state and $|s\>$ is the uniform
superposition (\ref{eq:uniform}).  We then evolve with the Hamiltonian
\be
  H = \Hfree - \beta |w\>\<w|
\label{eq:ham}
\ee
with parameters $\omega,\gamma$ to be determined in the analysis
below, for a time $T$ also determined below.  The goal is to choose
the parameters $\omega$ and $\gamma$ so that for some $T$ as small as
possible, the spatial component of the evolved state has a substantial
overlap on $|w\>$.

\section{Analysis of the algorithm}

To analyze the algorithm, we would like to determine the spectrum of
$H$ using our knowledge of the spectrum of $\Hfree$.  We do this using
the same techniques we applied to the Hamiltonian (\ref{eq:oldham}) in
\cite{CG03}.  An eigenvector of $H$, denoted $|\psi_a\>$, with
eigenvalue $E_a$, satisfies
\be
  H |\psi_a\> = (\Hfree - \beta |w\>\<w|) |\psi_a\> 
              = E_a |\psi_a\>
\,.
\label{eq:evdef}
\ee
Defining
\be
  \<w|\psi_a\> = \sqrt{R_a} |\phi_a\>
\label{eq:rdef}
\ee
where $|\phi_a\>$ is a normalized spin state, and $\sqrt{R_a}>0$ by
choice of phases, we can rewrite (\ref{eq:evdef}) as
\be
  (\Hfree - E_a)|\psi_a\> = \sqrt{R_a} \, \beta |\phi_a,w\>
\,.
\ee
Assuming $\Hfree-E_a$ is nonsingular, we can write the eigenstate of
$H$ as
\be
  |\psi_a\> = \frac{\sqrt R_a}{\Hfree - E_a} \beta|\phi_a,w\>
\,.
\label{eq:evec}
\ee
Consistency with (\ref{eq:rdef}) then gives the eigenvalue condition
\be
   |\phi_a\> = F(E_a) \beta |\phi_a\>
\ee
where
\be
  F(E) = \<w| \frac{1}{\Hfree - E} |w\>
\ee
operates on the spin degree of freedom.  To find eigenvalues of $H$,
we must look for values of $E$ such that the spin operator $F(E)
\beta$ has an eigenvector of eigenvalue 1.

In addition to finding eigenvalues of $H$, we need some facts about
its eigenvectors.  The normalization condition on $|\psi_a\>$ gives
\ba
  R_a^{-1} &= \<\phi_a,w|\beta\frac{1}{(\Hfree - E_a)^2}\beta|\phi_a,w\> \\
           &= \<\phi_a|\beta F'(E_a) \beta|\phi_a\>
\,.
\label{eq:r}
\ea
We also need the overlap of $|\psi_a\>$ with eigenvectors of $\Hfree$.
From (\ref{eq:evec}) we have
\be
  \<\E|\psi_a\> = \frac{\sqrt{R_a}}{\E - E_a} \<\E|\beta|\phi_a,w\>
\label{eq:evecoverlap}
\ee
where $|\E\>$ is an eigenvector of $\Hfree$ with eigenvalue $\E$.

For the free Hamiltonian (\ref{eq:hfree}), we find
\ba
  F(E) \beta &= \<w| \frac{\Hfree + E}{\Hfree^2 - E^2} |w\> \beta
                \label{eq:fcond} \\
             &= \frac{1}{N} \sum_\k \frac{\gamma \, c(\k) + \beta E}
                 {\E(\k)^2 - E^2} \label{eq:cancelodd} \\
             &= -\frac{\beta}{NE} + U(E) + \beta \, E \, V(E)
\,,
\ea
where in (\ref{eq:cancelodd}) we have canceled terms that are odd in
$\k$, and
\ba
  U(E) &= \frac{1}{N} \sum_{\k \ne 0} 
          \frac{\gamma \, c(\k)}{\E(\k)^2 - E^2} \label{eq:ue} \\
  V(E) &= \frac{1}{N} \sum_{\k \ne 0} 
          \frac{1}{\E(\k)^2 - E^2} \label{eq:ve}
\,.
\ea
If $|E| \ll \E(\k)$ for all $\k \ne 0$, then we can Taylor expand
$U(E)$ and $V(E)$ in powers of $E$.  We will need only the leading
order terms $U(0)$ and $V(0)$.  For large $N$, we have
\be
  U(0) \approx \frac{1}{(2\pi)^d} \int_{-\pi}^\pi 
               \frac{\gamma \, c(\k) \, \d^d k}{\E(\k)^2}
\,,
\ee
which is a convergent integral regardless of $d$.  For $d>2$ and $N$
large, we can also write $V(0)$ as a convergent integral,
\be
  V(0) \approx \frac{1}{(2 \pi)^d} \int_{-\pi}^\pi 
               \frac{\d^d k}{\E(\k)^2}
\,.
\ee
In $d=2$, this integral is logarithmically infrared divergent, and
instead we find
\be
  V(0) = \frac{1}{4\pi\omega^2} \ln N + O(1)
\,.
\ee

Now suppose we choose $\omega$ and $\gamma$ such that $U(0)=1$.  In
this case, we are simply looking for a zero eigenvalue of
$\beta(-\frac{1}{NE}+E \, V(0) )$.  We find such eigenvalues with
\be
  E_\pm \approx \pm \frac{1}{\sqrt{V(0) \, N}}
\,,
\ee
which indeed satisfy the condition $|E_\pm| \ll \E(\k)$ for all $\k
\ne 0$.  These eigenvalues are degenerate in the spin space, i.e., any
state $|\phi_\pm\>$ provides an eigenvector with the same eigenvalue.

\begin{figure}
\includegraphics[width=\columnwidth]{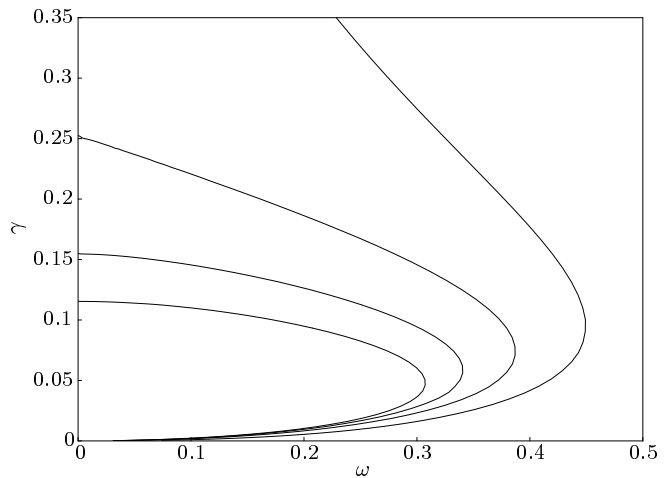}
\caption{Critical values of $(\omega,\gamma)$ for various dimensions.
From rightmost curve to leftmost curve, $d=2,3,4,5$.}
\label{fig:critical}
\end{figure}

The condition $U(0)=1$ can be satisfied by choosing
$u(\omega/\gamma)=\gamma$, where $u(\omega/\gamma)=\gamma \, U(0)$ is
a function only of $\omega/\gamma$.  Figure~\ref{fig:critical} shows
the critical curve in the $(\omega,\gamma)$ plane for $d=2$ through
$5$.  For any $\omega$ with $0<\omega<\omega^*$, where $\omega^*$ is
some dimension-dependent threshold value, there are two values of
$\gamma$ such that $U(0)=1$.  Note that with $\omega=0$, we recover
the results of \cite{CG03}.  Also, with $\gamma=0$, no solution of
$U(0)=1$ exists, so it was essential to include this additional term
(and the ability to fine tune its coefficient).

Having found the relevant eigenvalues, we need to determine the
corresponding eigenstates.  Using (\ref{eq:r}) we find
\be
  R_\pm^{-1} \approx \frac{1}{NE_\pm^2} + V(0) \approx 2 V(0)
\,,
\ee
and using (\ref{eq:evecoverlap}) we find
\be
  \<\eta,s|\psi_\pm\> = - \frac{\sqrt{R_\pm}}{E_\pm \sqrt{N}}
                        \<\eta|\beta|\phi_\pm\>
                \approx \mp \frac{1}{\sqrt 2}
\,,
\ee
where we have chosen the eigenstate of $H$ with $|\phi_\pm\> = \beta
|\eta\>$.  Therefore we have
\be
  |\eta,s\> \approx \frac{1}{\sqrt 2} (|\psi_-\> - |\psi_+\>)
\,,
\ee
and choosing $T = \pi / (2 |E_\pm|)$ produces the state
\be
  e^{-i H T} |\eta,s\> \approx \frac{1}{\sqrt 2} (|\psi_+\> + |\psi_-\>)
\ee
which has an overlap on $|\eta,w\>$ of $\sqrt{2R_\pm}$.

For $d>2$, we have shown that there is a $T=O(\sqrt N)$ that gives a
probability $O(1)$ of finding $w$.  For $d=2$, there is a $T=O(\sqrt{N
\log N})$ that gives an amplitude $O(1/\sqrt{\log N})$, so that
amplitude amplification \cite{BHMT00} can be used to find $w$ with a
probability $O(1)$ in time $O(\sqrt N \log N)$.

\section{Discussion}

We have described a continuous-time quantum walk algorithm for the
spatial search problem.  Using Dirac's insight of introducing spin to
take the square root in a relativistic dispersion relation, we have
found a Hamiltonian that locates a single marked item in the optimal
time of $O(\sqrt N)$ above the critical dimension ($d>2$), and that
runs in time $O(\sqrt N \log N)$ in $d=2$.

This algorithm is closely related to the discrete-time quantum walk
search algorithm of \cite{AKR04}.  Very similar techniques to the ones
we have used in this paper can also be applied to discrete-time
quantum walks \cite{CE03}.  This analysis for the algorithm of
\cite{AKR04} closely parallels the analysis above, which highlights
the similarity between the two kinds of algorithms.  However, there
are a few important differences.  The continuous-time algorithm
requires fine tuning the parameters $\omega$ and $\gamma$, whereas
there is (apparently) no equivalent fine tuning in the discrete-time
algorithm.  Also, the discrete-time algorithm has noticeably different
behavior depending on whether $N^{1/d}$ is odd or even, a difference
that is not seen in the continuous-time algorithm.  In short, although
the essential infrared features of the two kinds of algorithms are
identical, their detailed behaviors differ.

In high dimensions, our algorithm is very wasteful in terms of the
number of spin degrees of freedom: it uses a $2^{\lceil d/2
\rceil}$-dimensional spin space, whereas \cite{CG03} shows that no
spin degrees of freedom are required at all for $d>4$.  In comparison,
the discrete-time quantum walk search algorithm in \cite{AKR04} uses
$2d$ extra degrees of freedom.  The Dirac particle in $d$ dimensions
cannot be represented with fewer than $2^{\lceil d/2 \rceil}$ degrees
of freedom, but a continuous-time search algorithm with only $d+1$
degrees of freedom can arise from reproducing the Dirac algebra
(\ref{eq:anticom}) only on a subspace.  If the operators $\alpha_j$
and $\beta$ satisfy
\be
  \{\alpha_j,\alpha_k\} |\eta\> = 2 \delta_{j,k} |\eta\> \,, ~
  \{\alpha_j,\beta\} |\eta\> = 0 \,, ~
  \beta |\eta\> = |\eta\>
\label{eq:subanticom}
\ee
for some spin state $|\eta\>$, then the algorithm will work starting
from the state $|\eta,s\>$.  The condition (\ref{eq:subanticom}) is
sufficient to give $H_0^2 |\eta,\k\> = \E(\k)^2 |\eta,\k\>$.  The
previous analysis then shows that
\be
  |\psi_a\> = \frac{\sqrt{R_a}}{H_0 - E_a} |\eta,w\>
\ee
is an eigenstate of $H$ with eigenvalue $E_a$ provided
$-\frac{1}{NE_a}+U(E_a)+E_a V(E_a)=1$, where $U(E)$ and $V(E)$ are as
defined in equations (\ref{eq:ue}) and (\ref{eq:ve}).  The rest of the
analysis with two states $|\psi_\pm\>$ follows exactly as before.
Finally, we see that (\ref{eq:subanticom}) can be satisfied in a
$(d+1)$-dimensional spin space with basis $|0\>,|1\>,\ldots,|d\>$,
since in that case we can choose $\alpha_j = |0\>\<j| + |j\>\<0|$,
$\beta = 2|0\>\<0| - I$, and $|\eta\>=|0\>$.

Unlike the algorithm of \cite{CG03}, the algorithm of this paper
cannot be turned into an adiabatic algorithm.  With the Hamiltonian
(\ref{eq:oldham}), by starting in the state $|s\>$ and lowering the
parameter $\gamma$ from a large value to zero sufficiently slowly, the
adiabatic theorem guarantees that the system will remain near its
ground state, ending up close to the state $|w\>$.  In $d>4$, this can
be done in time $O(\sqrt N)$, and in $d=4$, it can be done in time
$O(\sqrt N \log^{3/2} N)$.  However, in the algorithm of the present
paper, states with $\k=0$ are not the ground state of the free
Hamiltonian (\ref{eq:hfree}); these states have zero energy, but this
is in the middle of the spectrum.  Although the adiabatic theorem
applies to any eigenstate, not just the ground state, states near the
middle of the spectrum of (\ref{eq:ham}) with $\omega,\gamma$ small have
very little overlap on $|w\>$, so that even perfectly adiabatic
evolution produces a state far from the desired one.

Finally, we note that the actual complexity of the spatial search
problem in $d=2$ is still an open question.  A gap of $\log N$ remains
between the best known algorithm and the lower bound of \cite{BBBV97}.
It would be interesting to improve the algorithm further or to show
that no such improvement is possible.

\acknowledgments

We thank Andris Ambainis, Wim van Dam, Edward Farhi, Julia Kempe, and
John Preskill for helpful discussions.
AMC received support from the Fannie and John Hertz Foundation, and JG
acknowledges the hospitality of the Caltech Institute for Quantum
Information, where this work was supported in part by the National
Science Foundation under grant EIA-0086038.  This work was also
supported in part by the Cambridge--MIT Institute, by the Department
of Energy under cooperative research agreement DE-FC02-94ER40818, and
by the National Security Agency and Advanced Research and Development
Activity under Army Research Office contract DAAD19-01-1-0656.


\bibliographystyle{apsrev_title}
\bibliography{main}

\end{document}